\documentstyle[12pt]{article}

\begin{document}
\vspace*{-2cm}
\hspace*{8cm}DEMIRM MEUDON 93/055
\hspace*{8.6cm}LPTHE PAR 93/56
\vskip 3cm
\centerline{ {\bf INFINITELY MANY STRINGS IN DE SITTER SPACETIME:}}
\vskip 5mm
\centerline{ {\bf EXPANDING AND OSCILLATING ELLIPTIC FUNCTION SOLUTIONS} }
\vskip 1.5cm
\centerline{{\bf H.J. de Vega}$^{(a)}$, {\bf A.L. Larsen}$^{(b,c)}$ {\bf and}
{\bf N. S\'{a}nchez}$^{(b)}$}
\vskip 1cm
\centerline{ {\bf December 1993}}
\vskip 4cm
\hspace*{-6mm}{\it (a)  Laboratoire de Physique
Th\'{e}orique et
Hautes Energies, Universit\'{e} Paris VI, Tour 16,
1er \'{e}tage, 4, Place Jussieu
75252 Paris, Cedex 05, France.}
\vskip 0.4cm
\hspace*{-6mm}{\it (b) Observatoire de Paris, Section
de Meudon, Demirm, Laboratoire
Associ\'{e} au CNRS UA 336, Observatoire de Meudon et \'{E}cole Normale
Sup\'{e}rieure, Paris. 5, Jules Janssen, 92195 Meudon Principal Cedex, France.}
\vskip  0.4cm
\hspace*{-6mm}{\it (c) On leave of absence
from Nordita,
Blegdamsvej 17, DK-2100 Copenhagen \O, Danemark.}
\newpage
\begin{centerline}
{\bf Abstract}
\bigskip
\end{centerline}
The exact general evolution of circular strings in $2+1$ dimensional
de Sitter spacetime is described closely and completely in terms of elliptic
functions. The evolution depends on a constant parameter $b$, related to the
string energy, and falls into three classes depending on whether $b<1/4$
(oscillatory motion), $b=1/4$ (degenerated, hyperbolic motion) or
$b>1/4$ (unbounded motion). The novel feature here is that one single
world-sheet generically describes {\it infinitely many} (different and
independent) strings. The world-sheet time $\tau$ is an infinite-valued
function of the string physical time, each branch yields a different
string. This has no analogue in flat spacetime. We compute the string
energy $E$ as a function of the string proper size $S$, and analyze it
for the expanding and oscillating strings. For expanding strings
$(\dot{S}>0)$: $E\neq 0$ even at $S=0$, $E$ decreases for small $S$ and
increases $\propto\hspace*{-1mm}S$ for large $S$. For an oscillating string
$(0\leq S\leq S_{max})$, the average energy $<E>$ over one oscillation
period is expressed as a function of $S_{max}$ as a complete elliptic
integral of the third kind.

For each $b$, the two independent solutions $S_+$ and $S_-$ are analyzed.
For $b<1/4$, all the strings of the $S_-$ solution are unstable
$(S_{max}=\infty)$ and never collapse to a point $(S_{min}\neq 0)$. $S_+$
describes one stable ($S_{max}$ is bounded) oscillating string and
$<E>$ is an increasing function of $b$ for $0\leq b\leq 1/4$. For $b>1/4,$ all
strings (for both $S_+$ and $S_-$) are unstable and have a collapse during
their evolution. For $b=1/4$, $S_-$ describes two strings (one stable and
one unstable for large de Sitter radius), while $S_+$ describes one
stable non-oscillating string.
\newpage
\section{Introduction and Results}
The study of string dynamics in curved spacetime reveales new insights and
new physical phenomena with respect to string propagation in flat spacetime
(and with respect to quantum fields in curved spacetime)
\cite{veg1,veg2,veg3}.
The results of
this programme are relevant both for fundamental (quantum) strings and for
cosmic strings, which behave essentially in a classical way.

Among the cosmological backgrounds of interest, de Sitter spacetime occupies
a special place. It is on one hand relevant for inflation, and on the
other hand, string propagation turns out to be specially interesting there.
String-instability, in the sense that the string proper length
grows indefinitely (proportional to the expansion factor of the universe) is
particularly present in de Sitter spacetime \cite{veg1,ven1,ven2}.

Recently, several progresses in the understanding of string propagation in
de Sitter spacetime have been performed [6-8]. The classical string
equations of motion (plus the string constraints) were shown to be
integrable in D-dimensional
de Sitter spacetime \cite{san1}.They are equivalent to a non-linear
sigma model on the grassmannian $SO(D,1)/O(D)$ with periodic boundary
conditions (for a closed string). In addition, the string constraints
imply a zero world-sheet energy-momentum tensor, and these constraints
are compatible
with the integrability. Moreover, the exact string dynamics in de Sitter
spacetime is equivalent to a generalized sh-Gordon model with a potential
unbounded from below \cite{san1}. The sh-Gordon function $\alpha(\sigma,\tau)$
has here a clear physical meaning: $H^{-1}\exp[\alpha(\sigma,\tau)/2]$
determines the proper size of the string ($H$ is the Hubble constant). In
$2+1$ dimensions, the string dynamics is exactly described by the
standard sh-Gordon equation.

More recently, a novel feature for strings in de Sitter spacetime
was found: Exact multi-string solutions
\cite{mic2}. Exact circular string solutions were found \cite{mic1}
describing two
different strings. One string is stable (the proper size is bounded),
and the other one is unstable (the proper size blows up) for large
de Sitter radius. Soliton methods, (the so-called "dressing method" in
soliton theory) were implemented using the linear problem (Lax pair) of
this system, in order to construct systematically exact string solutions
\cite{mic2}. The one-soliton string solution constructed in this way,
generically describe five different and independent strings: one stable
string and four unstable strings. These solutions (even the stable string)
do not oscillate in time. Exact string solutions oscillatory in
time in de Sitter universe were not found until now.

In this paper, we go further in the investigation of exact string solutions
in de Sitter spacetime. We find exact string solutions describing
{\it infinitely many} different and independent strings. The novel feature
here is that we have one single world-sheet but multiple (infinitely many)
strings. The world-sheet time $\tau$ turns out to be an infinite-valued
function of the target string time $X^0$ (which can be the hyperboloid time
$q^0$, the cosmic time $T$ or the static coordinate time $t$).
Each branch of $\tau$ as a
function of $q^0$ corresponds to a different string. In flat spacetime,
multiple string solutions are necessarily described my multiple
world-sheets. Here, a single world-sheet describes infinitely many different
and simultaneous strings as a consequence of the coupling with the
spacetime geometry. These strings do not interact among themselves; all the
interaction is with the curved spacetime.
\vskip 6pt
\hspace*{-6mm}We apply the circular string {\it Ansatz}:
\begin{eqnarray}
t=t(\tau),\;\;\;\theta=\sigma,\;\;\;r=\frac{1}{H}f(\tau)\nonumber
\end{eqnarray}
in $2+1$ dimensional de Sitter spacetime, particularly convenient in terms
of the static de Sitter coordinates $(t,r,\theta)$ (we also describe the
solutions in the hyperboloid and cosmic parametrizations). The string equations
of motion and constraints can be solved directly and completely in terms of
elliptic functions. They reduce to two decoupled first order differential
equations for
the time component $t(\tau)$ and the string radius $f(\tau)$:
\begin{eqnarray}
&\dot{t}=\frac{\sqrt{b}}{H(1-f^2)}&\nonumber
\end{eqnarray}
\begin{eqnarray}
&\dot{f}^2+V(f^2)=b;&\;\;\;\;V(f^2)=f^2-f^4\nonumber
\end{eqnarray}
The $f$-equation is solved by: $f^2(\tau)=\wp(\tau-\tau_o)+1/3$ where
$\wp$ is the Weierstrass elliptic function with discriminant
$\Delta=16b^2(1-4b)$, and $b$ and $\tau_o$ are integration constants
($\tau_o$ is generally complex and must be chosen so that $f(\tau)$ is
real for real $\tau$). The solutions depend on one constant parameter $b$
related to the string energy, and fall into three classes, depending on
whether $b<1/4\;(\Delta>0)$, $b=1/4\;(\Delta=0)$ or $b>1/4\;(\Delta<0)$.
As can be seen in the diagram $(f^2,V(f^2))$, Fig.1., in which the full string
dynamics takes place, these cases correspond to oscillatory motion and to
infinite (unbounded) motion.
\vskip 6pt
\hspace*{-6mm}The proper string size $S$ and energy $E$ of the circular
strings are given for all $f$ by:
\begin{eqnarray}
S=\frac{1}{H}f,\;\;\;\;E=\frac{1}{\alpha'H}\frac{(f\dot{f}-\sqrt{b})}
{f^2-1}.\nonumber
\end{eqnarray}
We find for an expanding string $(\dot{f}>0)$, see section 5:
\begin{eqnarray}
&E(f\approx 0)=\frac{\sqrt{b}}{\alpha'H}(1-f),&\nonumber
\end{eqnarray}
\begin{eqnarray}
&E(f=1)=\frac{1}{\alpha'H}\frac{(1+b)}{2\sqrt{b}},&\nonumber
\end{eqnarray}
\begin{eqnarray}
&E(f>>1)=\frac{1}{\alpha'H}f.&\nonumber
\end{eqnarray}
Notice also that the energy is non-zero, even at the collapse $(f=0)$, (except
for the degenerate case $f=b=0$, in which there is no string at all). It
follows from these expressions that for a string expanding
from zero radius,
the energy first decreases, and then increases for large $f$, proportional
to the invariant string size. For a string oscillating between $f=0$ and
$f_{max}=\sqrt{(1-\sqrt{1-4b})/2}\;$ (in the $b<1/4$-case, see Fig.1.), the
average energy $<E>$ over a period $T$ is:
\begin{eqnarray}
H\alpha'<E>=\frac{2\sqrt{b}}{T\sqrt{1-f_{max}^2}}\Pi(f^2_{max},
\frac{f_{max}}{\sqrt{1-f^2_{max}}}),\nonumber
\end{eqnarray}
in terms of the complete elliptic integral of third kind $\Pi$,\cite{gra}.

The string solutions in de Sitter spacetime enjoy conserved quantities
associated with the $O(3,1)$ rotations on the hyperboloid. For the circular
solutions under consideration here, the only non-zero component is given by:
\begin{eqnarray}
L_{01}=-L_{10}=2\pi\sqrt{b}.\nonumber
\end{eqnarray}
\vskip 6pt
\hspace*{-6mm}In the $b=1/4$-case, the Weierstrass elliptic function
degenerates into a hyperbolic function:
\begin{eqnarray}
f^2(\tau)=\frac{1}{2}[1+\sinh^{-2}(\frac{\tau-\tau_o}{\sqrt{2}})].\nonumber
\end{eqnarray}
Two real independent solutions appear for the choices $\tau_o=i\pi/2$ and
$\tau_o=0$, respectively:
\begin{eqnarray}
f^2_\pm(\tau)=\frac{1}{2}[\tanh(\frac{\tau}{\sqrt{2}})]^{\pm 2}.\nonumber
\end{eqnarray}
(They were previously found in ref. \cite{mic1})
We have also the solution $f^2_0=1/2$, corresponding to a stable string with
constant proper size $S_0=1/(\sqrt{2}H)$ (i.e., sh-Gordon function
$\alpha=0$). This solution was found in ref.\cite{mic1} and we do not discuss
it here.

The solution $f_-$ describes two different strings, I and II, as it can be
seen from the hyperboloid time $q^0_-(\tau)$, eq. (5.19), Fig.2a.
Here $\tau$ is a two-valued function of $q^0_-$: String I
corresponds to $-\infty<\tau<0$ and string II to $0<\tau<\infty$. The
proper size and energy $S_-$ and $E_-$ for both strings are given by
eq. (5.20). For $q^0_-\rightarrow\infty$, string I is unstable, while
string II is stable. Both $S_-(q^0_-\rightarrow\infty)$ and
$E_-(q^0_-\rightarrow\infty)$ blow up for string I (for which
$q^0_-\rightarrow\infty$ corresponds to $\tau\rightarrow 0_-$), while
they tend to constant values for string II (for which $q^0_-\rightarrow\infty$
corresponds to $\tau\rightarrow\infty$). String I starts with minimal size
$S_-=1/(\sqrt{2}H)$ and $E_-=1/(\alpha'H)$ at $\tau=-\infty$ and blows up
at $\tau=0$. String II starts with infinite size at $\tau=0$ but approaches
$S_-=1/(\sqrt{2}H)$ and $E_-=1/(\alpha'H)$ for $\tau\rightarrow\infty$.

The solution $f_+$ of this $b=1/4$-case describes only one stable string-
$q^0_-(\tau)$, eq. (5.26), is a monotonically increasing function of $\tau$-
with proper size $S_+$ and energy $E_+$ given by eqs. (5.26). The string
starts with $S_+=1/(\sqrt{2}H)$, $E_+=1/(\alpha'H)$ at $q^0_+=-\infty$,
it contracts until it collapses ($S_+=0$, $E_+=1/(2\alpha'H)$), then it
expands until it reaches the original size and energy for $q^0_+=
\infty$. The average energy is $<E_+>=1/(\alpha'H)$ which is equal to the
maximal energy. The string has minimal energy shortly after the collapse, as
follows from the general expression (5.11). For $b=1/4$ the evolution is
always
non-oscillatory. Even the stable string does not oscillate in time.
\vskip 6pt
\hspace*{-6mm}For $b<1/4$ there exist two real independent solutions
for the choices $\tau_o=0$ and $\tau_o=\omega'$, where
$\omega'$ is the
imaginary semi-period, eq. (4.13), of the Weierstrass function:
\begin{eqnarray}
&f^2_-(\tau)=\wp(\tau)+1/3,&\nonumber\\
&f^2_+(\tau)=\wp(\tau+\omega')+1/3.&\nonumber
\end{eqnarray}
$f_-$ and $f_+$ are oscillating solutions as functions of $\tau$. The
solution $f_-$ describes infinitely many strings; $f_-$ has infinitely
many branches $[0,2\omega],\;[2\omega,4\omega],\;...$, each of which
corresponds
to a different string ($\omega$ is the real semi-period, eq. (4.15), of the
Weierstrass function). This can be seen from the hyperboloid time
$q^0_-(\tau)$,
Fig.3a.: The world-sheet time $\tau$ is an infinite-valued function of
$q^0_-$. The hyperboloid time $q^0_-$ blows up at the boundaries of the
branches $\tau=\pm 2N\omega\;$ ($N$ being an integer):
\begin{eqnarray}
\mid q^0_-(\tau)\mid\sim\frac{1}{\mid 2N\omega-\tau\mid}.\nonumber
\end{eqnarray}
Further insight is obtained by considering the cosmic time $T_-$ and the
static coordinate time $t_-$. Closed expressions for them are given by
eqs. (5.33)-(5.35) and (5.40), in terms of Weierstrass $\zeta$ and
$\sigma$-functions, and also rewritten in terms of elliptic theta-functions.
The cosmic time $T_-$ is singular at $\tau=0,\;\tau=x/\mu,\;\tau=2\omega$
and similarly in the other branches. ($x,\;\mu$ are two real constants. $x$
is expressed as an incomplete elliptic integral of first kind while
$\mu=\sqrt{(1+\sqrt{1-4b})/2}\;$). The static coordinate time $t_-$, on the
other hand, is regular at the boundaries of the branches, but is singular
at $\mu\tau=2KN\pm x$:
\begin{eqnarray}
t_-(\tau)\sim\frac{1}{2\pi}\log\mid\mu\tau-2KN\mp x\mid,\nonumber
\end{eqnarray}
where $K$ is a complete elliptic integral of the first kind. It must be
noticed that although $f_-$ is periodic in $\tau$, the cosmic time is not,
i.e. $T_-(\tau)\neq T_-(\tau+2\omega)$, eq. (5.43). This implies that the
infinitely many strings are different (the difference in their invariant
proper size for a given cosmic time $T$ is given by eq. (5.48)), but they are
all of the same type: unstable. For instance, in the branch $\tau\in
[0,2\omega]$, the string starts at $\tau=0\;(q^0_-=-\infty)$ with infinite
size, then contracts to the minimal size $HS_-=\sqrt{(1+\sqrt{1-4b})/2}$
and eventually expands towards infinite size at $\tau=2\omega\;(q^0_-=
\infty)$. These solutions never collapse.

For the solution $f_+$ of the $b<1/4$-case, the string
dynamics takes place inside the
horizon. $f_+$, being a regularly oscillating function of $\tau$,
is then also a regularly oscillating function of the string times
$q^0_+,\;T_+$ and $t_+$. The static coordinate time $t_+$, from which one
easily deduces $q^0_+$ and $T_+$, is given in terms of theta-functions,
eq. (5.51), and reexpressed in terms of the Jacobi zeta-function, eq. (5.52).
The solution $f_+$ describes one stable string oscillating between its
minimal size $S_+=0$ (collapse) and its maximal size $HS_+=\sqrt{(1+
\sqrt{1-4b})/2}$. The string energy $E_+$ is given by eq. (5.56), and the
average energy $<E_+>$ is a monotonically increasing function of $b$,
for $b\in[0,1/4]$. It must be noticed that the string oscillations here do
not follow a pure harmonic motion as in flat Minkowski spacetime, but they
are precise superpositions of all frequencies $(2n-1)\Omega,\;(\Omega
=\pi\mu/(2K),\;n=1,2,...,\infty)$; here
the non-linearity of the string equations of motion
fixes the relation between the mode coefficients, and the basic frequency
$\Omega$ depends on the string energy.
\vskip 6pt
\hspace*{-6mm}For $b>1/4$ two real independent solutions are obtained for
$\tau_o=0$ and $\tau_o=\omega_2'$, where $\omega_2'$ is the imaginary
semi-period of the Weierstrass function:
\begin{eqnarray}
&f^2_-(\tau)=\wp(\tau)+1/3,&\nonumber\\
&f^2_+(\tau)=\wp(\tau+\omega_2')+1/3.\nonumber
\end{eqnarray}
In this case $f_-$ again describes infinitely many strings, all of them
are unstable. The difference with the $b<1/4$-case, is that here the strings
have a collapse during their evolution. For instance, in the branch
$\tau\in[0,2\omega_2]$, where $\omega_2$ is the real semi-period of the
Weierstrass function, the string starts with infinite size at
$\tau=0\;(q^0_-=-\infty)$, it then contracts until it collapses to a point
and then it expands towards infinite size again (at $\tau=2\omega_2\;
(q^0_-=\infty)$). In contrast to the $b<1/4$-case, the solution $f_+$ is
here just a time translated version of $f_-$:
\begin{eqnarray}
f^2_+(\tau)=f^2_-(\tau+\omega_2),\nonumber
\end{eqnarray}
and describes therefore essentially the same features as the solution $f_-$.
\vskip 6pt
\hspace*{-6mm}A summary picture of the main properties of the solutions
of this paper is
presented in Table I. The paper is organized as follows: In section 2 we
formulate the string dynamics in $2+1$ dimensional de Sitter spacetime
and apply to it the circular string {\it Ansatz}. In section 3 we describe
the problem in the static parametrization. In section 4 we find the closed
and complete string solutions in terms of elliptic functions. Section 5
deals with the physical interpretation of these solutions and the concept
of infinitely many strings.

\section{String Dynamics in 2+1 de Sitter Spacetime}
\setcounter{equation}{0}
It is well known that the $2+1$ dimensional de Sitter spacetime can be
considered as a $3$ dimensional hyperboloid embedded in 4 dimensional
flat Minkowski space:
\begin{equation}
ds^2=\frac{1}{H^2}\eta_{\mu\nu}dq^\mu dq^\nu,
\end{equation}
where $\mu =(0,1,2,3),\;\;\eta_{\mu\nu}=diag(-1,1,1,1)$, $H$ is the Hubble
constant and we require:
\begin{equation}
\eta_{\mu\nu}q^\mu q^\nu=1.
\end{equation}
The equations of motion for the bosonic string in the conformal gauge takes
the form \cite{mic1}:
\begin{equation}
q^\mu_{+-}-e^\alpha q^\mu=0,
\end{equation}
where:
\begin{equation}
e^{\alpha(\tau,\sigma)}\equiv -\eta_{\mu\nu}q^\mu_+ q^\nu_-,
\end{equation}
and we have introduced the notation $q^\mu_\pm=\frac{1}{2}(\partial_\sigma
\pm\partial_\tau)q^\mu,\;$ etc. The equations
of motion are as usual supplemented
by the constraints, that take the form:
\begin{equation}
\eta_{\mu\nu}q^\mu_\pm q^\nu_\pm=0.
\end{equation}

It was shown by de Vega and S\'{a}nchez \cite{san1} that the function $\alpha
(\tau,\sigma)$ fulfills the sh-Gordon equation:
\begin{equation}
\ddot{\alpha}-\alpha''-e^\alpha+e^{-\alpha}=0.
\end{equation}
Therefore, one can first look
for solutions
$\alpha (\tau,\sigma)$ to this equation
and then work backwards trying to solve the linear (in $q^\mu$) equation (2.3)
and finally impose the constraints (2.5). In the special case of circular
string configurations under consideration here it will turn out, however, that
the original equations of motion can be solved directly (and completely),
so in this case it is not necessary to first solve equation (2.6). Before
we come to the special solutions let us remark that $\alpha(\tau,\sigma)$
generally determines the invariant string size. This follows from the
observation that the line element (2.1) can be written as:
\begin{equation}
ds^2=\frac{1}{2H^2}e^{\alpha(\tau,\sigma)} (d\sigma^2-d\tau^2),
\end{equation}
i.e., we can identify:
\begin{equation}
S(\tau,\sigma)\equiv\frac{1}{\sqrt{2}H}e^{\alpha(\tau,\sigma)/2}
\end{equation}
as the invariant string size.

The circular strings are obtained by the following {\it Ansatz}:
\begin{equation}
q^\mu=(q^0(\tau),\;q^1(\tau),\;f(\tau)\cos\sigma,\;f(\tau)\sin\sigma).
\end{equation}
In this case the normalization condition (2.2) reads:
\begin{equation}
(q^0)^2-(q^1)^2-f^2=-1.
\end{equation}
The equations of motion (2.3) become:
\begin{equation}
\ddot{q}^0=e^\alpha q^0,
\end{equation}
\begin{equation}
\ddot{q}^1=e^\alpha q^1
\end{equation}
as well as:
\begin{equation}
\ddot{f}+f=e^\alpha f.
\end{equation}
The definition (2.4) is:
\begin{equation}
e^\alpha=(\dot{q}^0)^2-(\dot{q}^1)^2-\dot{f}^2+f^2,
\end{equation}
and the constraints (2.5) become:
\begin{equation}
(\dot{q}^0)^2-(\dot{q}^1)^2-\dot{f}^2-f^2=0.
\end{equation}
The 6 equations (2.10)-(2.15) give an overconstrained system for
$(q^0,q^1,f,\alpha)$. Due to the non-linear
nature it looks quite complicated,
but we will now see that the complete solution can be easily found.

Subtraction of (2.14) and (2.15) immediately yields:
\begin{equation}
e^\alpha=2f^2
\end{equation}
and then eq. (2.13) is integrated to:
\begin{equation}
\dot{f}^2+f^2-f^4={\rm const}\equiv b.
\end{equation}
This equation is generally
solved in terms of a Weierstrass elliptic function, and we
shall return to the explicit solution in section 4. Equation (2.10) is
formally solved by:
\begin{eqnarray}
&q^0=\sqrt{1-f^2}\sinh Ht,&\nonumber\\
&q^1=\sqrt{1-f^2}\cosh Ht,&
\end{eqnarray}
for $1-f^2\geq 0$ and by:
\begin{eqnarray}
&q^0=\sqrt{f^2-1}\cosh Ht,&\nonumber\\
&q^1=\sqrt{f^2-1}\sinh Ht,&
\end{eqnarray}
for $1-f^2\leq 0$. Here $t=t(\tau)$ and as usual we should make an extra
copy of $(q^0,q^1)$ to cover all of the de Sitter geometry (since
$q^0+q^1\geq 0$, using eqs. (2.18)-(2.19)). Now equation (2.15) is fulfilled
provided:
\begin{equation}
\dot{t}=\frac{\sqrt{b}}{H(1-f^2)}
\end{equation}
and (2.11),(2.12) are trivially fulfilled! The original system of equations
and constraints has now been reduced to the two separated first order equations
(2.17) and (2.20).

Let us for a moment return to the function $\alpha$
introduced in eq. (2.4). From
eq. (2.16) we find that $4f\dot{f}=e^\alpha\dot{\alpha},\;4f\ddot{f}=
(\ddot{\alpha}+\dot{\alpha}^2/2)e^\alpha$, so that eq. (2.17) leads to:
\begin{equation}
\ddot{\alpha}-e^\alpha+4be^{-\alpha}=0.
\end{equation}
The redefinitions $\tilde{\tau}=(4b)^{1/4}\tau,\;\alpha(\tau)=
\frac{1}{2}\log4b+\tilde{\alpha}(\tilde{\tau})$
yield:
\begin{equation}
\frac{d^2\tilde{\alpha}}{d\tilde{\tau}^2}-e^{\tilde{\alpha}}+
e^{-\tilde{\alpha}}=0,
\end{equation}
that is, the sh-Gordon equation, as was proved more generally by de Vega and
S\'{a}nchez \cite{san1}.
Note that in the special case where $b=1/4$ (corresponding
to $E=-2$ in the notation of Ref.\cite{mic1}) the redefinitions are trivial.
\section{Static Parametrization and String Radius}
\setcounter{equation}{0}
In this section we will show that the results of the previous section can
be obtained in an easier way by starting directly from
the static parametrization. In the
static parametrization the line element of $2+1$  dimensional de
Sitter spacetime takes
the form:
\begin{equation}
ds^2=-(1-H^2r^2)dt^2+\frac{dr^2}{(1-H^2r^2)}+r^2d\theta^2.
\end{equation}
Writing $x^\mu=(t,r,\theta)$ and
$g_{\mu\nu}=diag(-(1-H^2r^2),\;(1-H^2r^2)^{-1},
\;r^2)$ the equations of motion and constraints for the bosonic string in the
conformal gauge read $(\mu=0,1,2)$:
\begin{eqnarray}
&\ddot{x}^\mu-x''^\mu+\Gamma^\mu_{\rho\sigma}(\dot{x}^\rho\dot{x}^\sigma-
x'^\rho x'^\sigma)=0,&\nonumber\\
&g_{\mu\nu}\dot{x}^\mu x'^\nu=g_{\mu\nu}(\dot{x}^\mu\dot{x}^\nu+x'^\mu
x'^\nu)=0,&
\end{eqnarray}
where the non-vanishing components of the Christoffel symbol are:
\begin{eqnarray}
&\Gamma^r_{rr}=\frac{H^2r}{1-H^2r^2},\;\;\Gamma^r_{tt}=-H^2r(1-H^2r^2),&
\nonumber
\end{eqnarray}
\begin{eqnarray}
&\Gamma^r_{\theta\theta}=-r(1-H^2r^2),\;\;\Gamma^t_{rt}={{-H^2r}
\over{1-H^2r^2}},\;\;\Gamma^\theta_{\theta r}={{1}\over {r}}.&\nonumber
\end{eqnarray}
The {\it Ansatz} for the circular string in the static coordinates is:
\begin{equation}
t=t(\tau),\;\;\theta=\sigma,\;\;r=\frac{1}{H}f(\tau).
\end{equation}
In this case, equations (3.2) lead to:
\begin{eqnarray}
&\ddot{t}-2{{\dot{t}\dot{f}f}\over {1-f^2}}=0,&\nonumber
\end{eqnarray}
\begin{eqnarray}
&\ddot{f}+\frac{f\dot{f}^2}{1-f^2}-f(1-f^2)H^2\dot{t}^2+f(1-f^2)=0,&
\nonumber
\end{eqnarray}
\begin{eqnarray}
&-(1-f^2)H^2\dot{t}^2+{{\dot{f}^2}\over {1-f^2}}+f^2=0.&
\end{eqnarray}
The $\ddot{t}$-equation is immediately integrated and then all three equations
are consistently solved by:
\begin{eqnarray}
&\dot{f}^2+f^2-f^4=b,&\nonumber
\end{eqnarray}
\begin{eqnarray}
&\dot{t}={{\sqrt{b}}\over {H(1-f^2)}}.&
\end{eqnarray}
where $b$ is an integration constant. This is in agreement with the results of
section 2; compare with equations (2,17) and (2.20). Note that the constant
$b$ must be positive to ensure a real static coordinate time and that there
is a horizon at $f^2=1$.

We close this section by the following observation: The line
element in the static coordinates (3.1) is now:
\begin{equation}
ds^2=\frac{f^2}{H^2}(d\sigma^2-d\tau^2),
\end{equation}
that should be compared with eqs. (2.7)-(2.8), that is:
\begin{equation}
f^2=e^\alpha/2\equiv H^2S^2,
\end{equation}
which is in agreement with eq. (2.16).
\section{Expanding and Oscillating Elliptic Function Solutions}
\setcounter{equation}{0}
We now come to the explicit solution of equation (2.17) or equivalently of
the first equation of (3.5). The $\dot{t}$-equation will be analysed in
detail in the following section where we consider the physical interpretation
of the various solutions. Equation (2.17) is solved by:
\begin{equation}
f^2(\tau)=\wp(\tau-\tau_o)+1/3,
\end{equation}
where $\wp$ is the Weierstrass elliptic $\wp$-function \cite{gra}:
\begin{equation}
\dot{\wp}^2=4\wp^3-g_2\wp-g_3,
\end{equation}
with invariants:
\begin{equation}
g_2=4(\frac{1}{3}-b),\;\;\;g_3=\frac{4}{3}(\frac{2}{9}-b),
\end{equation}
discriminant:
\begin{equation}
\Delta\equiv g_2^3-27g_3^2=16b^2(1-4b),
\end{equation}
and roots $(e_1,e_2,e_3)$:
\begin{equation}
\frac{1}{6}(1+3\sqrt{1-4b}),\;\;\;\frac{1}{6}
(1-3\sqrt{1-4b}),\;\;\;-\frac{1}{3}.
\end{equation}
Finally
$\tau_o$ is a complex integration constant that must be carefully chosen
to obtain a real $f(\tau)$ for real $\tau$; recall that the Weierstrass
function is only real on a lattice in the complex plane.

It is convenient to write equation (2.17) in the form:
\begin{equation}
\dot{f}^2+V(f^2)=b;\;\;\;V(f^2)\equiv f^2-f^4
\end{equation}
and to consider separately the 3 cases $b=1/4\;(\Delta=0)$, $b<1/4\;(\Delta>0)$
and $b>1/4\;(\Delta<0)$, see Fig.1.
\vskip 12pt
\hspace*{-6mm}${\bf b=1/4}$: In this
case the Weierstrass function reduces to a hyperbolic
function and eq. (4.1) becomes:
\begin{equation}
f^2(\tau)=\frac{1}{2}(1+\sinh^{-2}(\frac{\tau-\tau_o}{\sqrt{2}})).
\end{equation}
Two real independent solutions are obtained by the choices $\tau_o=0$ and
$\tau_o=i\pi/\sqrt{2}$, respectively:
\begin{equation}
f^2_-(\tau)=\frac{1}{2}\tanh^{-2}\frac{\tau}{\sqrt{2}},
\end{equation}
\begin{equation}
f^2_+(\tau)=\frac{1}{2}\tanh^2\frac{\tau}{\sqrt{2}}.
\end{equation}
Note that:
\begin{eqnarray}
&f^2_-(-\infty)=\frac{1}{2},\;\;\;f^2_-(0)=\infty,\;\;\;
f^2_-(\infty)=\frac{1}{2},&
\nonumber\\
&f^2_+(-\infty)=\frac{1}{2},\;\;\;f^2_+(0)=0,\;\;\;
f^2_+(\infty)=\frac{1}{2}.&
\end{eqnarray}
These are the 2 solutions originally found by de Vega, S\'{a}nchez and
Mikhailov \cite{mic1}, corresponding to $\alpha>0$ and $\alpha<0$ respectively.
The interpretation of these solutions as a function
of the world-sheet time $\tau$ is clear from Fig.1: The solution (4.8)
expands from $f_-^2=1/2$ towards infinity and then contracts until it
reaches its original size. The solution (4.9) contracts from $f_+^2=1/2$
until it collapses. It then expands again until it reaches its
original size. The physical interpretation,
that is somewhat more involved, was described in
Ref\cite{mic1} and
will be shortly reviewed in section 5. There is actually also a
stationary solution for $b=1/4$, corresponding to $\alpha=0$, i.e.
a string sitting on the top
of the potential; see Fig.1. This solution with constant string size
$S=1/(\sqrt{2}H)$ was discussed in
Ref.\cite{mic1} and will not be considered here.
\vskip 12pt
\hspace*{-6mm}${\bf b<1/4}$: Here two real independent solutions are obtained
by the choices $\tau_o=0$ and $\tau_o=\omega'$, respectively:
\begin{equation}
f^2_-(\tau)=\wp(\tau)+1/3,
\end{equation}
\begin{equation}
f^2_+(\tau)=\wp(\tau+\omega')+1/3,
\end{equation}
where $\omega'$ is the imaginary semi-period of the Weierstrass function. It
is explicitly given by \cite{gra}:
\begin{equation}
\omega'=i\frac{\sqrt{2}K'(k)}{\sqrt{1+\sqrt{1-4b}}};\;\;\;k=\sqrt{\frac
{1-\sqrt{1-4b}}{1+\sqrt{1-4b}}}.
\end{equation}
Note that:
\begin{eqnarray}
&f^2_-(0)=\infty,\;\;\;f^2_-(\omega)=(1+\sqrt{1-4b})/2,\;\;\;f^2_-(2\omega)=
\infty,...&\nonumber\\
&f^2_+(0)=0,\;\;\;f^2_+(\omega)=(1-\sqrt{1-4b})/2,\;\;\;f^2_+(2\omega)=0,
...&
\end{eqnarray}
where $\omega$ is the real semi-period of the Weierstrass function:
\begin{equation}
\omega=\frac{\sqrt{2}K(k)}{\sqrt{1+\sqrt{1-4b}}},
\end{equation}
and $K'$ and $K$ are the complete elliptic integrals of first kind.
The interpretation of these solutions as a function of $\tau$
is clear from Fig.1: The solution (4.11) oscillates between infinity and its
minimal size $f_-^2=(1+\sqrt{1-4b})/2$ at the boundary of the potential, while
the solution (4.12) oscillates between $0$ and its maximal size
$f_+^2=(1-\sqrt{1-4b})/2$. The
physical interpretation will be considered in the following section.
\vskip 12pt
\hspace*{-6mm}${\bf b>1/4}$: In this last case two real independent solutions
are obtained
by the choices $\tau_o=0$ and $\tau_o=\omega_2'$, respectively:
\begin{equation}
f^2_-(\tau)=\wp(\tau)+1/3,
\end{equation}
\begin{equation}
f^2_+(\tau)=\wp(\tau+\omega_2')+1/3,
\end{equation}
where $\omega_2'$ takes the explicit form:
\begin{equation}
\omega_2'=i\frac{K'(\hat{k})}{b^{1/4}};\;\;\;\hat{k}
=\sqrt{\frac{1}{2}+\frac{1}{4\sqrt{b}}}.
\end{equation}
Note that:
\begin{eqnarray}
&f^2_-(0)=\infty,\;\;\;f^2_-(\omega_2)=0,\;\;\;f^2_-(2\omega_2)
=\infty,...&\nonumber\\
&f^2_+(0)=0,\;\;\;f^2_+(\omega_2)=\infty,\;\;\;f^2_+(2\omega_2)=0,...&
\end{eqnarray}
where:
\begin{equation}
\omega_2=\frac{K(\hat{k})}{b^{1/4}}.
\end{equation}
It should be stressed that in this case the
primitive semi-periods are $\hat{\omega}=(\omega_2-\omega'_2)/2$ and
$\hat{\omega}'=(\omega_2+\omega'_2)/2$, i.e.
$(2\hat{\omega},2\hat{\omega}')$ spans a
fundamental period
parallelogram in the complex plane.

The interpretation of the solutions (4.16)-(4.17)
as a function of $\tau$
follows from Fig.1: Both of them oscillates between zero size
(collapse) and infinite size (instability). The
physical interpretations will follow in the next section.
\section{Physical Interpretation. Infinitely many Strings}
\setcounter{equation}{0}
We now discuss the physical interpretation of the results obtained in section
4. For that purpose it is convenient to also describe the solutions in
terms of hyperboloid and/or comoving (cosmic) coordinates. The hyperboloid
coordinates were already introduced in section 2, and the relation to the
comoving coordinates $(T,X^1,X^2)$ is given by:
\begin{equation}
q^0=\sinh HT+\frac{H^2}{2}~e^{HT}~[(X^1)^2+(X^2)^2],
\end{equation}
\begin{equation}
q^1=\cosh HT-\frac{H^2}{2}~e^{HT}~[(X^1)^2+(X^2)^2],
\end{equation}
\begin{equation}
q^2=He^{HT}\,X^1,\;\;\;q^3=He^{HT}\,X^2,
\end{equation}
for $-\infty<T,X^1,X^2<+\infty$. That is:
\begin{equation}
T=\frac{1}{H}\log(q^0+q^1),\;\;\;X^1=\frac{1}{H}\frac{q^2}{(q^0+q^1)},\;\;\;
X^2=\frac{1}{H}\frac{q^3}{(q^0+q^1)}.
\end{equation}
The relation to the static
coordinates of sections 3 and 4 follows from eqs. (2.9) and (2.18)-(2.19).
The relation of the comoving coordinates to the static parametrization
(3.1) is given by:
\begin{equation}
T=\frac{\log\mid 1-H^2r^2\mid}{2H}+t,
\end{equation}
\begin{equation}
X^1=\frac{r\cos\theta}{\mid 1-H^2r^2\mid}e^{-Ht},
\end{equation}
\begin{equation}
X^2=\frac{r\sin\theta}{\mid 1-H^2r^2\mid}e^{-Ht},
\end{equation}
and the line element becomes:
\begin{equation}
ds^2=-dT^2+e^{2HT}[(dX^1)^2+(dX^2)^2],
\end{equation}
The string solution is then expressed in terms of the
comoving coordinates through equations (3.3),(4.1)
and (2.20) for $r$, $\theta=\sigma$ and $t$.

In describing the physical properties of our string solutions we will also
need the string energy that is computed from the spacetime string
energy-momentum tensor $(X^A=T,X^1,X^2)$:
\begin{equation}
\sqrt{-G}T^{AB}(X)=\frac{1}{2\pi\alpha'}\int d\sigma d\tau\,(\dot{X}^A
\dot{X}^B-X'^A X'^B)\,\delta^{(3)}(X-X(\tau,\sigma)).
\end{equation}
In the cases under consideration here, the cosmic
time $X^0=T\equiv T(\tau)$ is a function of $\tau$ only, and
the string energy becomes:
\begin{equation}
E(T)=\int d^2X\,\sqrt{-G}\;T^{00}(X)=\frac{1}{\alpha'}\frac{dT}{d\tau}.
\end{equation}
{F}rom this expression we can actually get a lot of information about the
energy
without using the explicit time evolution of the strings found in section 4.
{F}rom eqs. (5.5) and (3.5) it follows:
\begin{equation}
H\alpha' E=\frac{f\dot{f}-\sqrt{b}}{f^2-1};\;\;\;\;\dot{f}^2=b+f^2(f^2-1),
\end{equation}
giving the energy as a function of the invariant string size. We immediately
see that the energy is non-zero except for the degenerate case $f=b=0$, where
there is no string at all. At the horizon, the energy for an expanding string
$(\dot{f}>0)$ is:
\begin{equation}
H\alpha' E=\frac{b+1}{2\sqrt{b}};\;\;\;f=1
\end{equation}
while at $f=0$:
\begin{equation}
H\alpha' E=\sqrt{b};\;\;\;f=0
\end{equation}
Considering a string expanding from $f=0$ we find for small $f$:
\begin{equation}
H\alpha' E\approx\sqrt{b}(1-f),\;\;\;f\approx 0
\end{equation}
while for large $f$:
\begin{equation}
H\alpha' E\sim f;\;\;\;f>>1
\end{equation}
i.e. the string energy first decreases but eventually
increases proportionally to the invariant string size. Considering instead
a string configuration oscillating between $0$ and a maximal radius
$f_{max}\;(\dot{f}_{max}=0)$ we can calculate the average energy by
integrating over a period (say) $T$. The first term of eq. (5.11) does not
contribute since it is a total derivative of a periodic function
and from the second term we find:
\begin{eqnarray}
H\alpha' <E>\hspace*{-2mm}&=&\hspace*{-2mm}\frac{\sqrt{b}}
{T}\int_0^T\frac{d\tau}{1-f^2}\nonumber\\
\hspace*{-2mm}&=&\hspace*{-2mm}\frac{2\sqrt{b}}
{T}\int_0^{f_{max}}\frac{df}{\mid\dot{f}\mid(1-f^2)}
\nonumber\\
\hspace*{-2mm}&=&\hspace*{-2mm}\frac{2\sqrt{b}}
{T}\int_0^{f_{max}}\frac{df}{(1-f^2)\sqrt
{(f^2-f^2_{max})(f^2-(1-f^2_{max}))}}\nonumber\\
\hspace*{-2mm}&=&\hspace*{-2mm}\frac{2\sqrt{b}}
{T\sqrt{1-f^2_{max}}}\Pi(f^2_{max},\;f_{max}/\sqrt{1-f^2_{max}}),
\end{eqnarray}
where $\Pi$ is the complete
elliptic integral of third kind \cite{gra} and $f_{max}$
is the smallest root of $\dot{f}=0$:
\begin{equation}
f_{max}=\sqrt{(1-\sqrt{1-4b})/2}.
\end{equation}
Notice that this result
has been obtained without using any information about the
detailed time evolution of the string.

Let us finally remark that the string solutions in $2+1$ dimensional
de Sitter spacetime enjoy as conserved quantities those associated with
the $O(3,1)$ rotations on the hyperboloid (2.2):
\begin{equation}
L_{\mu\nu}=-L_{\nu\mu}=\int_0^{2\pi}d\sigma(q_\mu\dot{q}_\nu-q_\nu\dot{q}_\mu).
\end{equation}
\vskip 12pt
\hspace*{-6mm}After these introductory remarks let us
now reanalyse our string solutions in the three cases $b=1/4$, $b<1/4$
and $b>1/4$.
\subsection{The hyperbolic $b=1/4$ solutions}
This case was analysed in
Ref.\cite{mic1}, but let us restate and reinterpret some of the results here.
This is justified by the fact that many of the important features
of this case generalize to the $b\neq 1/4$ solutions (the elliptic solutions).
\vskip 12pt
\hspace*{-6mm}We first consider
the $f_-$-solution (4.8). The hyperboloid time $q^0(\tau)$
is obtained from equations (2.18), (2.19) and their
$(q^0+q^1\leq 0)$-counterparts, and by integrating eq. (2.20). The result is:
\begin{equation}
q^0_-(\tau)=\sinh\tau-\frac{1}{\sqrt{2}}\cosh\tau\coth\frac{\tau}{\sqrt{2}}.
\end{equation}
When we plot this function (Fig.2a.) we see that the string solution actually
describes 2 strings (I and II) \cite{mic1}, since
$\tau$ is a two-valued function of $q^0_-$.
For both strings the invariant size and energy
are given by eqs. (3.7) and (5.10),
respectively:
\begin{eqnarray}
S_-=\frac{1}{\sqrt{2}H}\coth\mid\frac{\tau}{\sqrt{2}}\mid,\nonumber
\end{eqnarray}
\begin{eqnarray}
E_-=\frac{1}{\alpha'H}\mid 1+{{1}\over{\cosh\sqrt{2}\tau-\frac{1}{\sqrt{2}}
\sinh\sqrt{2}\tau-1}}\mid,
\end{eqnarray}
but string I corresponds to $\tau\in\;]-\infty,0[\;$ and string II to
$\tau\in\;]0,\infty[\;$. Therefore, $q^0_-\rightarrow\infty$ corresponds to
$\tau\rightarrow 0_-$ for string I, but to $\tau\rightarrow\infty$ for string
II. More generally, when $q^0_-\rightarrow\infty$  both the invariant size
and the energy grow infinitely for string I, while they approach constant
values for string II. We conclude that string I is an unstable string
for $q^0\rightarrow\infty$, while string II is a stable string. More details
about the connection between hyperboloid time and world-sheet time for these
solutions can be found in Ref.\cite{mic1}.

Let us consider now the cosmic time, used to calculate the energy (5.20),
of the solution $f_-$ in a little more detail:
\begin{equation}
T_-(\tau)=\frac{1}{H}(\tau+\log\mid\frac{1}{\sqrt{2}}\coth\frac{\tau}{\sqrt{2}}
-1\mid).
\end{equation}
When we plot this function (Fig.2b.) we find that
$\tau$ is a three-valued function of $T_-$. What happens is that
the time interval $\tau\in\;]0,\infty[\;$ for
string II splits into two parts. These features
are easily understood when returning to the effective potential Fig.1.:
String I starts at $f_-^2=1/2$ for $\tau=HT_-=-\infty$, it then expands through
the horizon $f_-^2=1$ at:
\begin{equation}
\tau=-\sqrt{2}\log(1+\sqrt{2}),\;\;\;HT_-=\log 2-\sqrt{2}\log(1+\sqrt{2})
\end{equation}
and continues towards infinity for $\tau\rightarrow 0_-,\;HT_-
\rightarrow\infty$.
String II starts at infinity for $\tau=0_+,\;HT_-=\infty$ and contracts through
the horizon at:
\begin{equation}
\tau=\sqrt{2}\log(1+\sqrt{2}),\;\;\;HT_-=-\infty.
\end{equation}
This behaviour, approaching the horizon from the outside,
corresponds to the going backwards in cosmic time -part of Fig.2b. String II
then continues contracting from $f_-^2=1$ at:
\begin{equation}
\tau=\sqrt{2}\log(1+\sqrt{2}),\;\;\;HT_-=-\infty,
\end{equation}
until it collapses at $\tau=HT_-=\infty$.
\vskip 12pt
\hspace*{-6mm} We now consider briefly the $f_+$-solution (4.9). In this case
the hyperboloid time is given by:
\begin{equation}
q^0_+(\tau)=\sinh\tau-\frac{1}{\sqrt{2}}\cosh\tau\tanh\frac{\tau}{\sqrt{2}},
\end{equation}
which
is a monotonically increasing function of $\tau$. The $f_+$-solution therefore
describes only one string. The proper size and energy are
given by:
\begin{eqnarray}
S_+=\frac{1}{\sqrt{2}H}\tanh\mid\frac{\tau}{\sqrt{2}}\mid,\nonumber
\end{eqnarray}
\begin{eqnarray}
E_+=\frac{1}{\alpha'H}(1-{{1}\over{\cosh\sqrt{2}\tau-\frac{1}{\sqrt{2}}
\sinh\sqrt{2}\tau+1}}).
\end{eqnarray}
We can also express this solution in terms of the cosmic time:
\begin{equation}
T_+(\tau)=\frac{1}{H}
\left[\tau+\log(1-\frac{1}{\sqrt{2}}\tanh\frac{\tau}{\sqrt{2}})\right],
\end{equation}
but since everything now takes place well inside the horizon this will not
really give us more insight. The string starts with $f_+^2=1/2$ for $\tau=
HT_+=-\infty$ with the energy $E_+=1/(\alpha'H)$. It then contracts until
it collapses for $\tau=HT_+=0$ where the energy is reduced to
$E_+=1/(2\alpha'H)$. It now expands
again and eventually reaches $f_+^2=1/2$ for
$\tau=HT_+=\infty$, where it has regained its original energy. Note
that the string has the minimal energy for $\tau=\frac{1}{\sqrt{2}}
\log(\sqrt{2}+1),$ i.e. shortly after expanding from $f=0$,
thus confirming the general result obtained
from equation (5.11). We also remark that in this
case the average energy is actually equal to the maximal energy
$<E_+>=1/(\alpha'H)$.
\vskip 12pt
\hspace*{-6mm}Let us close
this analysis of the $b=1/4$ case by
mentioning that for both the $f_+$ and the $f_-$-solutions there is only
one non-vanishing component of $L_{\mu\nu}$ introduced in (5.18). It is
explicitly given by:
\begin{equation}
L_{10}=\pi.
\end{equation}
\subsection{The elliptic $b<1/4$ solutions}
The calculations are here going to be somewhat
more complicated since we deal with elliptic functions as compared to
hyperbolic functions. It is therefore convenient to first introduce a more
compact notation. Defining:
\begin{equation}
\mu\equiv\sqrt{\frac{1+\sqrt{1-4b}}{2}},\;\;\;\nu\equiv
\sqrt{\frac{1-\sqrt{1-4b}}{2}},
\end{equation}
so that $\mu^2+\nu^2=1,\;\;0\leq\nu\leq 1/\sqrt{2}\leq\mu\leq 1,\;\;\mu\nu=
\sqrt{b},$ we find from eqs. (4.13),(4.15):
\begin{equation}
k=\frac{\nu}{\mu},\;\;\;\omega=\frac{K(k)}{\mu},\;\;\;\omega'=
i\frac{K'(k)}{\mu}.
\end{equation}
The solutions (4.11) and (4.12) can be written as:
\begin{equation}
f^2_-(\tau)=\frac{\mu^2}{{\mbox{sn}}^2[\mu\tau\mid k]},
\end{equation}
\begin{equation}
f^2_+(\tau)=\nu^2 {\mbox{sn}}^2[\mu\tau\mid k],
\end{equation}
respectively.
\vskip 12pt
\hspace*{-6mm}Consider first
the $f_-$-solution (5.31). It is clear from eq. (4.14) and the
periodicity in general that we have infinitely many branches $[0,2\omega],\;
[2\omega,4\omega],\;...$. We will see in a moment that each of these
branches actually corresponds to one string, that is, the $f_-$-solution
describes
infinitely many strings. For that purpose we will need the
hyperboloid time and the
cosmic time as a function of $\tau$. Both of them are expressed in terms of
the static coordinate time $t$, that is obtained by integrating (2.20):
\begin{equation}
Ht_-(\tau)=\zeta(x/\mu)\tau+
\frac{1}{2}\log\mid\frac{\sigma(\tau-x/\mu)}{\sigma(\tau+x/\mu)}\mid,
\end{equation}
where $\zeta$ and $\sigma$ are the Weierstrass $\zeta$ and $\sigma$-functions
and $x$ is a real constant obeying $\mbox{sn}[x\mid k]=\mu$, i.e. $x$ is
expressed
as an incomplete elliptic integral of the first kind. The expression (5.33) can
be further rewritten in terms of theta-functions:
\begin{equation}
Ht_-(\tau)=\frac{\mu\tau\pi}{2K}\frac{\vartheta'_1}
{\vartheta_1}(\frac{\pi x}{2K})+
\frac{1}{2}\log\mid\frac{\vartheta_1(\frac{\pi(\mu\tau-x)}{2K})}
{\vartheta_1(\frac{\pi(\mu\tau+x)}{2K})}\mid,
\end{equation}
and finally as:
\begin{eqnarray}
Ht_-(\tau)=&\hspace*{-2mm}\frac{1}{2}\log\mid{{\sin\left[{{\pi}\over
{2K}}(\mu\tau-x)\right]}\over{\sin\left[{{\pi}\over{2K}}(\mu\tau+x)\right]}}
\mid+{{\mu\tau\pi}\over{2K}}
{{\vartheta'_1}\over{\vartheta_1}}
({{\pi x}\over{2K}})&\nonumber
\end{eqnarray}
\begin{eqnarray}
&-2\sum_{n=1}^\infty{{q^{2n}}\over{n(1-q^{2n})}}
\sin(\frac{n\pi\mu\tau}{K})\sin(\frac{n\pi x}{K}),&
\end{eqnarray}
where $q=e^{-\pi K'/K}$.
In the latter expression we have isolated all the real
singularities in the first term. To be more specific we see that the static
coordinate time is singular for $\mu\tau\rightarrow 2KN\pm x$, where
$N$ is an integer, with the
asymptotic behaviour:
\begin{equation}
t_-(\tau)\rightarrow\pm\frac{1}{2H}\log\mid\mu\tau -2KN\mp x\mid;\;\;\;\;
\tau\rightarrow \frac{2K}{\mu}N\pm\frac{x}{\mu}.
\end{equation}
On the other hand $t_-(\tau)$ is completely regular at the boundaries of
the branches, i.e. for $\tau=0,\;\pm2\omega,\;\pm4\omega,... $. These results
can be easily translated to the hyperboloid time and cosmic time obtained from
eqs.
(2.18),(2.19) and (5.5), respectively.
We find the explicit form of the hyperboloid time $q^0_-(\tau)$ from eqs.
(2.18)-(2.19),(5.31) and (5.34):
\begin{equation}
q^0_-(\tau)=-\frac{\Omega\vartheta'_1(0)}{2\pi}~\frac{e^{\Omega\tau\frac
{\vartheta'_1}{\vartheta_1}(\Omega y)}\vartheta_1(\Omega(y-\tau))+
e^{-\Omega\tau\frac
{\vartheta'_1}{\vartheta_1}(\Omega y)}\vartheta_1(\Omega(y+\tau))}
{\vartheta_1(\Omega\tau)\vartheta_1(\Omega y)},
\end{equation}
where:
\begin{equation}
\Omega\equiv\frac{\pi\mu}{2K},\;\;\;\;y\equiv\frac{x}{\mu}.
\end{equation}
Notice that the singularities (5.36) that originated by the zeroes of
$\vartheta_1(\Omega(y\pm\tau))$ cancel in $q^0_-(\tau)$ so that
$q^0_-(2\omega N\pm y)$ is finite.
$q^0_-(\tau)$ blows up for $\tau=0,\;\pm2\omega,\;
\pm4\omega,...$, like:
\begin{equation}
\mid q^0_-\mid\propto\mid\frac{1}{2N\omega-\tau}\mid,
\end{equation}
where $N$ is again an integer.
This demonstrates that the world-sheet time $\tau$
is actually an infinite valued
function of $q^0_-$, and that the solution $f_-$ therefore describes
{\it infinitely
many strings} (see Fig.3a). This should be compared with the $b=1/4$
case where
we found a solution describing two strings. In that case
the two strings were of completely different type and had completely different
physical interpretations. In the present case we find infinitely many strings
but they are all of the same type. In the branch $\tau\in[0,2\omega]$ (say)
the string starts with infinite string size at $\tau=0,\;\;q^0_-=-\infty$. It
then contracts to its minimal size $f_-^2=(1+\sqrt{1-4b})/2$ and reexpands
towards infinity at $\tau=2\omega,\;\;q^0_-=\infty$. This solution, and the
infinitely many others of the same type, are unstable strings.
\vskip 12pt
\hspace*{-6mm}The cosmic time of the $f_-$-solution is obtained by
combining eqs. (5.5),(5.31) and (5.35). Expanding also the $\log$-term of
eq. (5.5) in terms of $q$ we find:
\begin{eqnarray}
&HT_-(\tau)=\Omega\,\tau\frac{\vartheta'_1}{\vartheta_1}(\Omega y)+
\log\mid{{\Omega\vartheta_1(\Omega(\tau-y))\vartheta'_1(0)}\over
{\pi\vartheta_1(\Omega\tau)\vartheta_1(\Omega y)}}\mid &\nonumber
\end{eqnarray}
\begin{eqnarray}
&=\log\mid\frac{\Omega\sin(\Omega
(\tau-y))}{\sin(\Omega\tau)}\mid-\log\mid\frac{\pi\vartheta_1(\Omega y)}
{\vartheta'_1(0)}\mid+\Omega\tau
\frac{\vartheta'_1}{\vartheta_1}(\Omega y)&\nonumber
\end{eqnarray}
\begin{eqnarray}
&-4\sum_{m=1}^\infty\frac{q^{2m}}
{m(1-q^{2m})}
\sin(m\Omega y)\sin(m\Omega(2\tau-y)).&
\end{eqnarray}
It can be shown that $0\leq\Omega\leq 1$ and $1.246..\leq y\leq\pi/2$. It
follows that:
\begin{equation}
0<y<\omega<2\omega-y<2\omega.
\end{equation}
The cosmic time (5.40) is singular at $\tau=0,\;y,\;2\omega$ but regular at
$\tau=2\omega-y$ and similarly in the other branches. The singularity of the
static coordinate time (5.35) at $\tau=2\omega-y$ has been canceled by
adding the $\log$-term of eq. (5.5), see Fig.3b. Therefore, the interpretation
of the string solution in the branch $\tau\in[0,2\omega]$ (say), as seen in
comoving coordinates, is as follows: The string starts with infinite
size and energy at $\tau=0,\;\;HT_-=\infty$. It then contracts and passes the
horizon from the outside at $\tau=y,\;\;HT_-=-\infty$. The string now
continues contracting from the inside of the horizon at $\tau=y,\;\;
HT_-=-\infty$ until it reaches the minimal size at:
\begin{equation}
\tau=\omega,\;\;\;\;HT_-=\frac{\pi}{2}\frac{\vartheta'_1}{\vartheta_1}(\Omega
y)
+\frac{1}{2}\log\frac{1-\sqrt{1-4b}}{2}.
\end{equation}
The energy
is here given by $H\alpha'E_-(\tau=\omega)=2\sqrt{b}/(1-\sqrt{1-4b})$. From now
on the string expands again. It passes the horizon from the inside after
finite cosmic time and continues towards infinity for $HT_-\rightarrow\infty$.

It is an interesting observation that the cosmic time is not periodic in
$\tau$, i.e. $T_-(\tau)\neq T_-(\tau+2\omega)$, although the string size is.
Explicitly we find:
\begin{equation}
T_-(\tau+2\omega)-T_-(\tau)=\frac{\pi}{H}
\frac{\vartheta'_1}{\vartheta_1}(\Omega y).
\end{equation}
This means that $f_-(\tau)$ really describes infinitely many strings
with different invariant size
at a given cosmic time. To be more specific let us
consider a fixed cosmic time $T_-$ and the corresponding string times:
\begin{equation}
T_-\equiv T_-(\tau_1)=T_-(\tau_2)=...,
\end{equation}
where $\tau_1\in[0,2\omega[,\;\tau_2\in[2\omega,4\omega[...$. Taking for
simplicity a cosmic time $HT_->>1$ we have (see Fig.3b.):
\begin{equation}
\tau_n=\frac{n\pi}{\Omega}+\epsilon_n,\;\;\;\;\epsilon_n<<1.
\end{equation}
To the lowest orders we find from eq. (5.40):
\begin{equation}
HT_-=-\log\epsilon_n+n\pi\frac{\vartheta'_1}{\vartheta_1}(\Omega y)+
{\cal O}(\epsilon_n),
\end{equation}
so that:
\begin{equation}
\epsilon_n=\exp[-HT_-+n\pi\frac{\vartheta'_1}{\vartheta_1}(\Omega y)+...]
\end{equation}
The invariant string sizes are then:
\begin{equation}
HS_-(\tau_n)=f_-(\tau_n)\approx\frac{1}{\epsilon_n}=\exp[HT_--n\pi
\frac{\vartheta'_1}{\vartheta_1}(\Omega y)+...]
\end{equation}
i.e. they are separated by a multiplicative factor. This expression of
course is only valid as long as $\epsilon_n<<1$, so $n$ should not be
too large.
\vskip 12pt
\hspace*{-6mm}We now consider
the $f_+$-solution (5.32). In this case the dynamics takes
place well inside the horizon. The possible singularities of the
hyperboloid time $q^0_+$ and the cosmic time $T_+$ therefore coincide with
the singularities of the static coordinate time $t_+$. The static coordinate
time is again obtained from eq. (2.20) which we first rewrite as:
\begin{equation}
H\dot{t}_+(\tau)=\frac{\sqrt{b}}{2/3-\wp(\tau+\omega')}=
\sqrt{b}\;\left[1+\frac{b}{\wp(\tau)-(b-1/3)}\right].
\end{equation}
Integration leads to:
\begin{equation}
Ht_+(\tau)=\tau(\sqrt{b}+\zeta(a))+\frac{1}{2}\log\mid\frac{\sigma(\tau-a)}
{\sigma(\tau+a)}\mid,
\end{equation}
where $a$ is a complex constant obeying $\wp(a)=b-1/3,\;\;$ i.e.
$\mbox{sn}[a\mu\mid k]=1/\nu$. It
follows that $a\mu=iK+x$ where $x$ is real and
$\mbox{sn}[x\mid k]=\mu$. Again we can express the static coordinate time in
terms of theta-functions:
\begin{equation}
Ht_+(\tau)=\tau[\sqrt{b}+\frac{\mu\pi}{2K}\frac{\vartheta'_4}{\vartheta_4}
(\frac{\pi x}{2K})]+\frac{1}{2}
\log\mid\frac{\vartheta_4(\frac{\pi(\mu\tau-x)}{2K})}
{\vartheta_4(\frac{\pi(\mu\tau+x)}{2K})}\mid,
\end{equation}
or in terms of the Jacobi zeta-function $zn$ \cite{gra}:
\begin{equation}
Ht_+(\tau)=\tau(\sqrt{b}+\mu~ zn(x,k))-2\sum_{n=1}^\infty
\frac{q^n}{n(1-q^{2n})}\sin(\frac{n\pi\mu\tau}{K})\sin(\frac{n\pi x}{K}),
\end{equation}
where $q=e^{-\pi K'/K}$. In this form we see explicitly that $t_+$ consists
of a linear term plus oscillating terms.
The cosmic time takes the form (see eqs. (3.13) and (5.5)):
\begin{eqnarray}
&HT_+(\tau)=\frac{1}{2}\log(1-\nu^2{\mbox{sn}}^2
[\mu\tau\mid k])+Ht_+(\tau)&\nonumber
\end{eqnarray}
\begin{eqnarray}
&=\tau[\sqrt{b}+\Omega\frac{\vartheta'_4}{\vartheta_4}(\Omega y)]+
\log\mid\frac{\Omega\vartheta_4(\Omega(\tau-y))\vartheta'_1(0)}
{\pi\vartheta_1(\Omega y)\vartheta_4(\Omega\tau)}\mid. &
\end{eqnarray}
Notice that the argument of the $\log$ has no real zeroes:
\begin{eqnarray}
&HT_+(\tau)=\tau[\sqrt{b}+\Omega\frac{\vartheta'_4}{\vartheta_4}(\Omega y)]+
\log\mid\frac{\Omega\vartheta'_1(0)}{\pi\vartheta_1(\Omega y)}\mid &\nonumber
\end{eqnarray}
\begin{eqnarray}
&-4\sum_{n=1}^{\infty}\frac{q^n}{n(1-q^{2n})}\sin(n\pi\mu\Omega)
\sin(n\pi\Omega(2\tau-y)).&
\end{eqnarray}
The static coordinate
time and the cosmic time are therefore completely regular functions
of $\tau$, and it follows that
the string solution $f_+$, which is {\it oscillating regularly}
as a function of
world-sheet
time $\tau$, is also oscillating regularly when expressed in terms of
hyperboloid time or cosmic time. This solution represents one {\it stable}
string.

Series expansion analogous to eq. (5.54) hold for the string radius:
\begin{equation}
f_+(\tau)=\nu ~\mbox{sn}[\mu\tau\mid k]=\frac{2\pi}{K(k)\sqrt{1+k^2}}
\sum_{n=1}^{\infty}\frac{q^{n-1/2}}{1-q^{2n-1}}\sin((2n-1)\Omega\tau).
\end{equation}
We see from eq. (5.54)-(5.55) that the string oscillations do not follow a
pure harmonic motion as in flat Minkowski spacetime, but they are
precise superpositions of all frequencies $(2n-1)\Omega$ ($n=1,2,...,\infty$)
with uniquely defined coefficients. The non-linearity of the string equations
in de Sitter spacetime fixes the relation between the mode coefficients.
In the present case the basic frequency $\Omega$ depends on the
string energy, while in Minkowski spacetime the frequencies are
merely $n$.

Using eqs. (5.5) and (5.32) the energy (5.11) of this solution is given by:
\begin{eqnarray}
E_+\hspace*{-2mm}&=&\hspace*{-2mm}\frac{1}{\alpha'}
\frac{d}{d\tau}(\frac{1}{2H}\log(1-\nu^2
{\mbox{sn}}^2[\mu\tau\mid k])+t_+(\tau))\nonumber\\
\hspace*{-2mm}&=&\hspace*{-2mm}\frac{1}{H\alpha'}
[\sqrt{b}+\mu~ zn(x,k)]\;+\;"oscillating\;terms",
\end{eqnarray}
and by averaging over a period $2K/\mu$, the average energy $<E_+>$ is just
the square bracket term. Let us consider this energy in a little more detail.
In the limit $b\rightarrow 0$, where the amplitude of the string size
oscillation goes to $0$ (see Fig.1.), we find:
\begin{equation}
k\rightarrow 0,\;\;\;\mu\rightarrow 1,\;\;\;x\rightarrow\pi/2,
\end{equation}
and then $<E_+>\rightarrow 0$. This is not surprising since in the
limit $b\rightarrow 0$ the $f_+$-solution ceases to exist. The other
interesting limit is $b\rightarrow 1/4$ where the elliptic solutions turn
into hyperbolic solutions. We find:
\begin{equation}
k\rightarrow 1,\;\;\;\mu\rightarrow 1/\sqrt{2},\;\;\;\tanh x\rightarrow
1/\sqrt{2},
\end{equation}
and then $<E_+>\rightarrow 1/(H\alpha')$, in agreement with eq. (5.26). More
generally a numerical analysis shows that $<E_+>$ is a
monotonically increasing function of
$b$ for $b\in [0,1/4]$. The energy (5.56) should be compared with the
average energy computed directly from eq. (5.16). In
the special case here eq. (5.16) gives:
\begin{equation}
H\alpha'<E_+>=\sqrt{b}~\frac{\Pi(\nu^2,k)}{K(k)},
\end{equation}
which is indeed equivalent to the square bracket term of (5.56) \cite{abr}.
\vskip 12pt
\hspace*{-6mm}Finally we remark that the only non-vanishing component of
$L_{\mu\nu}$ introduced in (5.18) is again $L_{10}$. This holds for both
the $f_-$ and the $f_+$-solutions:
\begin{equation}
L_{10}=2\pi\sqrt{b}.
\end{equation}
This expression is actually general for all the solutions under
consideration here since it is obtained directly from the
equations
of motion (3.5). The result (5.28) is therefore just the special case $b=1/4$,
and (5.60) is also valid for $b>1/4$, that will be considered in the following
subsection.
\subsection{The elliptic $b>1/4$ solutions}
Before analysing the solutions let us also in this case introduce an
alternative notation for the elliptic functions. We first write:
\begin{equation}
\omega_2=\frac{K(\hat{k})}{\alpha},\;\;\;\;\omega_2'=
\frac{iK'(\hat{k})}{\alpha};\;\;\;\;\alpha\equiv
b^{1/4},
\end{equation}
and $\hat{k}$ is given by (4.18). The solutions (4.16) and (4.17) are then
expressed in terms of Jacobi elliptic functions:
\begin{equation}
f^2_-(\tau)=\alpha^2\frac{1+\mbox{cn}[2\alpha\tau\mid \hat{k}]}
{1-\mbox{cn}[2\alpha\tau\mid \hat{k}]},
\end{equation}
\begin{equation}
f^2_+(\tau)=\alpha^2\frac{1-\mbox{cn}[2\alpha\tau\mid \hat{k}]}
{1+\mbox{cn}[2\alpha\tau\mid \hat{k}]},
\end{equation}
respectively.
\vskip 12pt
\hspace*{-6mm}Consider first the $f_-$-solution (5.62). The static
coordinate time, obtained by integrating (2.20), is:
\begin{equation}
Ht_-(\tau)=\zeta(x/\alpha)\tau+\frac{1}{2}\log\mid
\frac{\sigma(\tau-x/\alpha)}
{\sigma(\tau+x/\alpha)}\mid.
\end{equation}
This expression is formally identical to (5.33), but remember that the
parameters of the Weierstrass functions are different now. The real
constant $x$ is here obeying:
\begin{equation}
\mbox{cn}[2x\mid\hat{k}]=\frac{1-\alpha^2}{1+\alpha^2}.
\end{equation}
In terms of theta-functions, following the same steps as in subsection 5.2,
$t_-(\tau)$ can be written in the same form as (5.35), but with
$q^2=-e^{-\pi K'(\hat{k})/K(\hat{k})}$.
The analysis of this function and of the corresponding
hyperboloid time and cosmic time is now similar to the analysis in the
$b<1/4$-case
after equation (5.35), so we shall skip the details. The result is that the
$f_-$-solution describes infinitely many strings of the same type, in the
same sense as in subsection 5.2. The only difference is that in the present
case
the strings have a collapse during their evolution: Considering the
branch $\tau\in[0,2\omega_2]$ (say) the string starts with infinite string
size at $\tau=0,\;q^0_-=-\infty$. It then contracts until it collapses, after
which it expands again and reaches infinite string size at $\tau=2\omega_2,\;
q^0_-=\infty$. This is again an unstable string.
\vskip 12pt
\hspace*{-6mm}The physical interpretation of the $f_+$-solution easily
follows from the fact that:
\begin{equation}
f^2_+(\tau)=f^2_-(\tau+\omega_2),
\end{equation}
i.e. it is just a time translated version of $f_-$.
\section{Conclusion}
We have found the exact general evolution of circular strings in $2+1$
dimensional de Sitter spacetime. We have expressed it closely and completely
in terms of elliptic functions. The solution generically describes
infinitely many (different and independent) strings, and depends on one
constant parameter $b$ related to the string energy. A summary of the
main features and conclusions of our results is presented in Table I.

\newpage
\begin{centerline}
{\bf Table I}
\end{centerline}
\vskip 30pt
\hspace*{-6mm}Circular string evolution in de Sitter spacetime. For each
$b,$ there exists two independent solutions $f_-$ and $f_+$:
\vskip 30pt
\begin{tabular}{|l|l|l|}\hline
$ $&  $ $ & $ $ \\
$\hspace{6mm}b$ & $\hspace{27mm}f_-$ & $\hspace{28mm}f_+$\\
$ $ & $ $ & $ $ \\ \hline
$ $ & $ $ & $ $ \\
$ $ & Infinitely many different strings. & One stable oscillating string  \\
$ $ & All are unstable ($f_{-}^{max}=\infty$), & ($0\leq f_+\leq f_{+}^{max}$,
where $f_{+}^{max}=$ \\
$b<1/4$ &  and never collapse to a point &
$\sqrt{(1-\sqrt{1-4b})/2}\;).\;<E_+>$ is \\
$ $ & ($f_{-}^{min}=\sqrt{(1+\sqrt{1-4b})/2}\;>0$) &
a bounded monotonically increa-\\
$ $ &  $ $ & sing function of $b\in[0,1/4]$
\\
$ $ & $ $ & $ $ \\ \hline
$ $ & $ $ & $ $ \\
$ $ & Infinitely many strings. All of & Infinitely many strings similar \\
$b>1/4$ & them are unstable $(f_{-}^{max}=\infty)$ & to $f_-$. In this case
$f_+$ is just a \\
$ $ & and they collapse to a point & time-translation of $f_-$: \\
$ $ & ($f_{-}^{min}=0$) & $f_+(\tau)=f_-(\tau+\omega_2)$ \\
$ $ & $ $ & $ $ \\ \hline
 $ $ & $ $ &  $ $ \\
$ $ & Two different and non-oscillating & One non-oscillating and stable\\
$b=1/4$ & strings $f_-^{(I)}$ and $f_-^{(II)}$. $f_-^{(I)}$ is &
string. $f_{+}^{min}=0$, $E_{+}^{min}=\frac{1}{2H\alpha'}$ \\
$ $ & unstable and $f_-^{(II)}$ is stable for &
$f_{+}^{max}=\frac{1}{\sqrt{2}H}$,
$E_{+}^{max}=\frac{1}{H\alpha'}$ \\
$ $ & large de Sitter radius. & $ $ \\
 $ $ & $ $ &  $ $ \\ \hline
\end{tabular}
\newpage
\begin{center}
{\bf Figure Captions}
\end{center}
\vskip 12pt
Fig.1. The potential $V(f^2)=f^2-f^4$ defined in equation (4.6). For
$b\leq 1/4$ it
acts effectively as a barrier. The horizon corresponds to $f^2=1$.
\vskip 36pt
\hspace*{-6mm}Fig.2a. The hyperboloid time $q_-^0$ in the $b=1/4$ case,
given by equation
(5.19), as a function of $\tau$. Notice that $\tau$ is a two-valued
function of $q_-^0$.
\vskip 36pt
\hspace*{-6mm}Fig.2b. The cosmic time $T_-$ in the $b=1/4$ case,
given by equation (5.21), as
a function of $\tau$. $T_-$ is singular for $\tau=0$ and
$\tau=\sqrt{2}\log(1+\sqrt{2})$, and thus $\tau$ is a three-valued function of
$T_-$.
\vskip 36pt
\hspace*{-6mm}Fig.3a. The hyperboloid time $q^0_-$ as a function of $\tau$
in the elliptic case $b<1/4$. Each of the infinitely many branches corresponds
to one string.
\vskip 36pt
\hspace*{-6mm}Fig.3b. The cosmic time $T_-$ in the $b<1/4$ case,
given by equation (5.40), as
a function of $\tau$. Notice that $T_-$ is {\it not} periodic,
as explained in subsection 5.2.
\end{document}